\documentclass[12pt]{article}
\usepackage{amssymb,amsmath,graphicx,mathtools}
\makeatletter
\usepackage{empheq}

\usepackage{geometry}
\makeatletter

\@addtoreset{equation}{section}
\def\section{\@startsection {section}{1}{\z@}{-2.5ex plus -1ex minus
 -.2ex}{1.3ex plus .2ex}{\large\bf}}
\def\subsection{\@startsection{subsection}{2}{\z@}{-2.25ex plus%
 -1ex minus -.2ex}{0.5ex plus .2ex}{\bf}}

\def\rpr{r'}
\def\all{\mu}

\def\bett{\nu}

\def\cd{\!\cdot\!}

\def\bp{{\mbox{\boldmath $p$}}}

\def\bq{{\mbox{\boldmath $q$}}}

\def\bm{{\mbox{\boldmath $m$}}}

\def\bpm{\begin{pmatrix}}
\def\epm{\end{pmatrix}}
  
\newcommand{\cg}{\mathfrak{g}}

\newcommand{\RR}{\mathbb{R}}
\newcommand{\CC}{\mathbb{C}}
\newcommand{\tr}{{\rm tr}}
\newcommand{\ad}{{\rm ad}}
\newcommand{\Ad}{{\rm adj}}
\newcommand{\id}{{\rm id}}
\def\bee{\begin{equation}}
\def\eee{\end{equation}}
\def\bea{\begin{eqnarray}}
\def\eea{\end{eqnarray}}
\newtheorem{theorem}{Theorem}[section]

\newtheorem{lemma}[theorem]{Lemma}

\newtheorem{definition}[theorem]{Definition}

\begin{document}
\parskip 4pt
\parindent 8pt

\begin{center}

\baselineskip 24 pt

{\Large \bf Classical $r$-matrices for the generalised \\ Chern-Simons formulation of  3d gravity }

\baselineskip 20 pt

\vspace{.2cm}

{ \bf Prince~K.~Osei } \\
Perimeter Institute,  Waterloo, Ontario, 
Canada \\
{posei@perimeterinstitute.ca}

\vspace{.2cm}

 {\bf  Bernd~J.~Schroers}   \\
Department of Mathematics and Maxwell Institute for Mathematical Sciences \\
 Heriot-Watt University, 
Edinburgh EH14 4AS, United Kingdom \\ 
{b.j.schroers@hw.ac.uk}

\vspace{0.3cm}

{20 February 2018}
\baselineskip 16 pt

\end{center}

\begin{abstract}
\noindent 
We study the conditions for classical $r$-matrices to be compatible with the generalised Chern-Simons action for 3d  gravity. Compatibility means  solving the  classical Yang-Baxter equations with a prescribed symmetric part for  each of the real Lie algebras and bilinear pairings  arising  in the generalised  Chern-Simons action.
We give a new construction of $r$-matrices via a generalised complexification and  derive  a non-linear set of   matrix  equations determining the  most general compatible $r$-matrix.  We  exhibit new   families of  solutions and show that they contain some known $r$-matrices for special parameter values. 
\end{abstract}


\section{Introduction and Motivation}

The application of the combinatorial quantisation programme \cite{AGSI,AGSII,AS,Schroers,BNR,MeusburgerSchroers1, MeusburgerSchroers2,MN}  to  the Chern-Simons formulation of 3d gravity \cite{AT,Witten,Carlipbook} has provided a systematic way of studying  the  role of quantum groups and non-commutative geometry in 3d quantum gravity.  What emerges in this framework is that  quantisation deforms the  classical phase space  geometry into a non-commutative geometry in which the  model spacetimes are replaced by non-commutative spaces and the local  isometry groups (gauge groups for the Chern-Simons theory)  by quantum isometry groups.

In the combinatorial quantisation procedure, one  begins with the  description of the Poisson structure on the  classical phase space  in terms of a classical $r$-matrix. In its original formulation for  generic Chern-Simons theories this description  is due to Fock and Rosly \cite{FockRosly}. It requires that the $r$-matrix satisfies a certain compatibility condition with the inner product used in defining the Chern-Simons action. When  the  $r$-matrix is the   classical limit of a quantum $R$-matrix associated to  a certain quantum group,  one can  formulate the quantum theory in terms of the  representation theory of that quantum group, see  \cite{AS} for the general framework and also \cite{sissatalk,Cracow} for survey accounts  related to 3d gravity.
 
In the case of the Chern-Simons formulation of 3d gravity,  $r$-matrices satisfying the compatibility condition are known for all signatures and values of the cosmological constant \cite{MSkappa}.   The associated quantum groups  are deformations of the classical isometry groups and thus natural candidates for the `quantum isometry groups'  of 3d quantum gravity. 

Unfortunately, Fock and Rosly's compatibility requirement  does not specify the  classical $r$-matrix uniquely. 
For a given classical theory, several compatible  $r$-matrices  are possible in general, and the associated  quantum groups are equally valid contenders for the role of  `quantum isometry group'.

The  family of classical $r$-matrices  and corresponding quantum groups that can be associated to the  most general Chern-Simons action for 3d gravity  via the Fock-Rosly compatibility requirement  has not yet been fully characterised.  
The quantum groups are  known to include a family of Drinfeld  quantum doubles and Maijd's bicrossproduct quantum groups with spacelike, timelike and null deformation parameters \cite{MSkappa,OseiSchroers1}. However, recent work \cite{BHM} shows that there are more and  that, at the infinitesimal Lie bialgebra level, all the known solutions can in fact be viewed as classical doubles.
There are also partial results on the relationship between different compatible $r$-matrices \cite{OseiSchroers1}.

Our goal here is to formulate the general equations which determine compatible $r$-matrices in 3d gravity, and to exhibit some new solutions.  It turns out to be mathematically natural to consider a  generalised form of the  Chern-Simons action for  3d gravity,  based on the most  general symmetric bilinear   form for the  relevant Lie algebra.  This generalised action was previously  considered 
in  \cite{Witten,MielkeBaekler,BonzomLivine,MSkappa}. In addition to the  usual  gravitational  term it contains  a non-standard term which has  been   interpreted as an analogue of the Immirzi term in  4d  gravity \cite{BonzomLivine}, though this interpretation has recently been questioned \cite{AGKY}.

The family of real Lie algebras used in the Chern-Simons formulation of 3d gravity depends on the speed of light and the value of the cosmological constant. It includes $so(p,q)$ with $p+q=4$ as well as the Poincar\'e and Euclidean group in, respectively, 2+1 and 3 dimensions. All of these are much studied in geometry and physics. Our results are therefore also of interest in the context of the bialgebra structures for these Lie algebras  and of  the Poisson-Lie algebra structures of the corresponding groups.
They generalise previous studies of $r$-matrices for the Euclidean and Poincar\'e group in  \cite{Stachura} and  \cite{OS2}\footnote{Readers referring to \cite{OS2} should be aware of the `dual' use of the parameter $\lambda$ in that paper compared to the current paper.}.

We should stress that, in general, the Chern-Simons formulation of 3d gravity is not equivalent to the Einstein or metric formulation of 3d gravity. The differences arise because the frame field is necessarily non-degenerate in the metric formulation but may degenerate in the Chern-Simons formulation, and because of different notions of gauge invariance in the two formulations. The resulting  phase spaces may therefore, in general, be different. This issue  has  been discussed extensively in the literature, starting with the papers  \cite{Mess,Messcomment} (showing that the resulting phases spaces may be the same under certain assumptions)  and \cite{Matschull}(exhibiting an example where they differ). This discussion has some parallels with that comparing gauge  and metric formulations of gravity in 4d, see,  e.g.,  \cite{TM}, but the flatness  of classical solutions in the Chern-Simons formulation leads to special features in 3d. This paper is not intended as a contribution to this discussion. We take the Chern-Simons formulation as the starting point of our treatment, and explore its consequences with a view to gaining a systematic understanding of its quantisation.

Our paper is organised as follows.
 We start in Sect.~2 with a review of the generalised Chern-Simons action for 3d gravity, the  description  of the relevant Lie algebras as generalised complexifications of $so(3)$ and $so(2,1)$. We give  a precise formulation of the compatibility requirement for an $r$-matrix in this context, and explain the reality condition we impose.  
  In Sect.~3  we derive  a particular class of solutions  by a process of (generalised) complexification of the standard $r$-matrix for the Lie bialgebra of $sl(2,\RR)\simeq so(2,1)$.  Sect.~4  contains the main results of this paper. We derive a set of non-linear coupled equations which characterises the most general compatible $r$-matrix in 3d gravity. In Sect.~5 we derive two  new families of solutions, and show how they relate to the  complexified solutions of Sect.~3 and that they include some known solutions as special cases. Sect.~6 contains our conclusions. 
 
\section{Lie algebras, Chern-Simons actions and compatible $r$-matrices}
\subsection{Lie algebra conventions}
We use the  conventions of \cite{OS2},  so  write  $\cg$ for either $so(3)$ or $so(2,1)$, with generators $J_a$,  $a=0,1,2$.   The metric $\eta_{ab}=\eta^{ab}$    is, respectively,   the Euclidean metric diag$(1,1,1)$ or the Lorentzian metric diag$(1,-1,-1)$. The Lie brackets of $\cg$ are  then
\bee
\label{so123}
[J_a,J_b]=\epsilon_{abc}J^c,
\eee
where indices are raised  with $\eta^{ab}$ and  we  adopt the convention $\epsilon_{012}=\epsilon^{012}=1$. 
We make frequent use of the Killing form as  an invariant (possibly Lorentzian) inner product   $\langle \cdot, \cdot \rangle$ on $\cg$, and assume that it is normalised so that 
\bee
\label{innerprod}
\langle J_a,  J_b\rangle =\eta_{ab}.
\eee
We write $A^t$ for the transpose of a map $A: \cg\rightarrow \cg$ with respect to the inner product  \eqref{innerprod}, i.e.,
\bee
\label{transposemap}
\langle v,Aw\rangle = \langle A^t v,w\rangle \qquad \forall v,w \in \cg. 
\eee
When we expand  elements of $\cg$ via
\bee
p = p^a J_a, \quad q=  q^a J_a,
\eee
we write $\bp=(p^0,p^1,p^2)$  for the coordinate vector and 
\bee
\langle p,q \rangle = \bp\cd \bq. 
\eee
We also use the following  notation  for the dual of a vector as a linear form:
\bee
\label{transposevector}
v^t = \langle v, \cdot \rangle, \quad v \in \cg. 
\eee

We write $\cg_\lambda$  for the family of Lie algebras which arise as isometry Lie algebras in 3d gravity. Depending on the real  parameter $\lambda$ (which is related to the cosmological constant, see \cite{Cracow} and our discussion in Sect.~\ref{cssection} ),  $\cg_\lambda$ is the Lie algebra  of the  Poincar\'e, de Sitter or anti-de Sitter  group or their Euclidean analogues in three dimensions, with brackets
\bee
\label{isoalg}
[J_a,J_b]=\epsilon_{abc}J^c, \quad  [J_a,P_b]=\epsilon_{abc}P^c, \quad [P_a,P_b]=\lambda \epsilon_{abc}J^c.
\eee

The Lie algebra $\cg_\lambda$ can  be interpreted as the real form of a generalised complexification of $\cg$, and  this viewpoint will prove useful in the following. We refer the reader to \cite{Meusburger} and \cite{MSquart} for details but briefly summarise the main results.
The idea is to introduce  a formal parameter $\theta$ satisfying 
\bee
\theta^2 =\lambda,
\eee
and to set 
\bee
P_a=\theta J_a.
\eee
Then  the commutators  \eqref{isoalg} follow  from \eqref{so123} by extending the Lie brackets linearly in $\theta$. More formally,  one defines the  ring  $R_\lambda =(\RR^2,+,\cdot )$ as  the  commutative ring obtained from $\RR^2$ with the usual addition  and the $\lambda$-dependent multiplication law 
\begin{equation}
 (a+\theta b)\cdot (c + \theta d)=(ac+\lambda bd) + \theta (ad+bc)\quad \forall a,b,c,d\in \RR,
\end{equation}
and defines  $\cg_\lambda$ as the realification of $\cg \otimes R_\lambda$ \cite{Meusburger}.

The Lie algebra $\cg_\lambda$ has a two-parameter family of non-degenerate symmetric Ad-invariant  bilinear   forms \cite{Witten}. Defining 
\begin{align}
\label{pair} &t( J_a,J_b)=0, & &t(P_a,P_b)=0, &
&t( J_a,P_b)=\eta_{ab},  \\
\label{othform} &s(J_a,J_b)=\eta_{ab}, & &s(J_a,P_b)=0, &
&s(P_a,P_b)=\lambda\eta_{ab},
\end{align}
the most general such form is given by 
\bee
\label{inprod}
(\cdot, \cdot)_\tau=  \alpha t(\cdot,\cdot)+ \beta s(\cdot,\cdot),
\eee
in terms of two real  parameters $\alpha,\beta$.  It turns  out \cite{MSkappa} that the condition for the non-degeneracy for \eqref{inprod} can conveniently be written in terms of the complexified   parameter $\tau = \alpha +\theta \beta\in R_\lambda$ as 
\bee
\label{taucond}
\tau
\bar\tau= \alpha^2-\lambda \beta^2 \neq 0. 
\eee

\subsection{ Chern-Simons theory and 3d gravity}
\label{cssection}
The Chern-Simons theory on a 3d manifold depends on a gauge group and an invariant, non-degenerate symmetric  bilinear form on the Lie algebra of that gauge group. One recovers  the Einstein-Hilbert action for 3d gravity for any signature and value of the cosmological constant from the Chern-Simons action by choosing the appropriate local isometry group $G_\lambda$  with Lie algebra $\cg_\lambda$ as gauge group and using the non-degenerate form \eqref{pair}, see \cite{AT,Witten}. Here we are interested in  the Chern-Simons action with the more general  bilinear form  \eqref{inprod}. This was previously  considered in \cite{Witten,MielkeBaekler,BonzomLivine} and, in our notation, in  \cite{MSkappa}
 
Consider a three-dimensional spacetime manifold $M^3$ of  the product topology $\RR\times S,$ where $S$ is an oriented two-dimensional manifold, possibly with handles and punctures. Physically, $S$ represents 'space' and the punctures particles.  The gauge field of the Chern-Simons theory is locally a one-form $A$ on the spacetime with values in the Lie algebra $\cg_{\lambda}$. It can be expanded in  terms of the generators $J_{a}$ and $P_{a}$ as
\begin{equation}
A=\omega_{a}J^{a}+e_{a}P^{a},
\end{equation}
 where $\omega=\omega^{a}J_{a}$ is geometrically interpreted as the spin connection on the frame bundle and the one-form $e_a$ as a dreibein. The curvature of this connection combines the Riemann curvature $R$, the torsion $T$ and a cosmological term, see  \cite{MSkappa} for details. 

The Chern-Simons action for the gauge field $A$ is 
\begin{equation}
\label{CSaction}
I_{\tau}(A)=\int_{M}\left(A\wedge dA\right)_{\tau}+\frac{1}{3}(A\wedge[A,A])_{\tau}.
\end{equation}
Integrating by parts and ignoring boundary terms, this can be expanded as  
 \begin{align}
\label{C-S action2}
I_{\tau}(A)&=\alpha\int_{M}\left(2e^{a}\wedge R_{a}+\frac{\lambda}{3}\epsilon_{abc}e^{a}\wedge e^{b}\wedge e^{c}\right) \nonumber\\
&+\beta\int_{M}\left(\omega^{a}\wedge d\omega_{a}+\frac{1}{3}\epsilon_{abc}\omega^{a}\wedge\omega^{b}\wedge\omega^{c}+\lambda e^{a}\wedge T_{a}\right). 
\end{align}
 The first term is the usual Einstein-Hilbert action for 3d gravity with a cosmological
constant. 
To make contact with  the physical constants of 3d gravity  identify
\begin{equation}
\alpha = \frac {1}{16\pi G}, \qquad  \lambda =-c^2\Lambda,
\end{equation}
where $G$ is Newton's constant in 2+1 dimensions, $c$ is the speed of light and taken to be imaginary in the case of Euclidean signature,   and $\Lambda$ is the cosmological constant. 

The term  proportional to $\beta$ contains the Chern-Simons action for the spin connection and a cosmological contribution. There have been attempts to interpret this term  as   the analogue of the Immirzi term in $4d$ \cite{BonzomLivine}, but this is contentious \cite{AGKY}.

A discussion of the classical equations of motion obtained when varying the action \eqref{C-S action2} with respect to $e_a$ and $\omega_a$, treated as independent variables, can be found in \cite{MSkappa}.

\subsection{Bi-algebras and classical $r$-matrices}
We refer the reader to  \cite{CP,Majidbook} for details on the background reviewed in this short section. 
A Lie bialgebra $(\cg,[\;,\;],\delta)$ is a Lie algebra $(\cg,[\;,\;])$  equipped with a map $\delta:\cg\mapsto \cg\otimes  \cg$ (the
cocommutator, or cobracket) satisfying the following condition:
\begin{description}
 \item{(i)} $\delta:\cg\mapsto \cg\otimes \cg$ is a skew-symmetric linear map, i.e., $\delta:\cg\mapsto \wedge^2\cg$
\item{(ii)} $\delta$ satisfies the coJacobi identity  $
(\delta\otimes \text{id})\circ \delta(X)+\mbox{cyclic}=0, \quad\forall X\in \cg. $
\item{(iii)} For all $X,Y\in \cg$, \;  $  \delta([X,Y])=(\ad_X\otimes 1+1\otimes \ad_X)\delta(Y)-(\ad_Y\otimes 1+1\otimes \ad_Y)\delta(X).$
\end{description}
An element $r\in\wedge^{2}\cg$ is said to
be a coboundary structure of the Lie bialgebra $(\cg,[\;,\;],\delta)$ if
 $ \delta(X)=\ad_X(r)=[X\otimes1+1\otimes X,r]$. 

For any Lie algebra $\cg,$ let $r=r^{ab}X_a\otimes Y_{b}\in \cg\otimes \cg$,  $r_{21}=\sigma(r)=r^{ab}Y_b\otimes X_a$ and set 
\bee r_{12}=r^{ab}X_a\otimes Y_{b}\otimes1,\;r_{13}=r^{ab}X_a\otimes1\otimes Y_{b},\;r_{23}=r^{ab}1\otimes X_a\otimes Y_{b}.
\eee
  The classical Yang-Baxter map is  the map
 \begin{equation}
\mbox{CYB}:\cg^{\otimes2}\rightarrow \cg^{\otimes3},\;\;r\mapsto[[r,r]]=[r_{12,}r_{13}]+[r_{12,}r_{23}]+[r_{13,}r_{23}].
\end{equation}
It is easy to check that CYB restricts to a map $\wedge^{2}\cg\rightarrow\wedge^{3}\cg.$
The equation \begin{equation}
\label{CYBE} [[r,r]]=0,
\end{equation} is called the classical Yang-Baxter equation(CYBE). Any solution $r\in \cg\otimes \cg$ of the CYBE  is called a classical $r$-matrix.
If $[[r,r]]$ is non-zero but an invariant element of $\cg\otimes \cg\otimes \cg$ then $r$ is said to satisfy the modified classical Yang-Baxter equation(MCYBE).
The triple
$(\cg,[\;,\;],r)$ defines a coboundary Lie bialgebra if and only if 
\bee \ad_X([[r,r]])=0,\quad \ad_X(r+r_{21})=0,\quad \forall X\in \cg,
\eee
where $\ad$ is the usual Lie algebra adjoint action extended to products in the usual way (as a derivation).

\subsection{Compatible $r$-matrices}

As reviewed above, a Chern-Simons theory requires a gauge group and an  invariant, non-degenerate, symmetric bilinear form on the Lie algebra  of the  gauge group. In the Fock-Rosly construction, a  classical $r$-matrix is said to be compatible with a Chern-Simons action if it satisfies the CYBE (\ref{CYBE}) and its symmetric part is equal to the Casimir associated  to the invariant, non-degenerate symmetric bilinear form used in the Chern-Simons action. Fock and Rosly went on to show how to describe the Poisson structure of an extended phase space for the Chern-Simons theory in terms of a compatible $r$-matrix.

Decomposing a compatible  $r$-matrix into  the symmetric Casimir  part $K$ and antisymmetric  part $\rpr $   via $r =K+\rpr $ we have the identity \cite{CP}
\begin{equation}
[[K+\rpr ,K+\rpr ]]=[[K,K]]+[[\rpr , \rpr ]].
\end{equation}  
Therefore, an $r$-matrix is compatible with a Chern-Simons action if its symmetric part equals the associated Casimir and its antisymmetric part satisfies the modified classical Yang-Baxter equation (MCYBE)
\bee 
\label{MCYBE}
 [[\rpr , \rpr ]]=- [[K,K]].
\eee

We can now apply this prescription to the Chern-Simons action \eqref{CSaction}.
The  Casimir  in $\cg_\lambda$  associated to the bilinear invariant form \eqref{inprod} is 
\bee
\label{Ktau}
K_\tau = \frac{\alpha }{\tau\bar \tau}(J_a\otimes P^a+P_a\otimes J^a)
-\frac{\beta}{\tau\bar\tau}(\lambda J_a\otimes J^a + P_a\otimes P^a).
\eee
It is  shown in  \cite{MSkappa}\footnote{In the corresponding expression  in \cite{MSkappa}, there is a missing factor of 2 in the second term. This term was not considered further  there, so the missing factor does not affect the conclusions of that paper.}  that 
\begin{align}
\label{omegatau}
\Omega_\tau &=[[K_\tau,K_\tau]] \nonumber \\
& \all \epsilon_{abc}(\lambda J^a\otimes J^b\otimes J^c+J^a\otimes P^b\otimes P^c
+P^a\otimes J^b\otimes P^c+ P^a\otimes P^b\otimes J^c)\nonumber \\
&+\bett\epsilon_{abc}( P^a\otimes P^b\otimes P^c+\lambda P^a\otimes J^b\otimes J^c
+\lambda J^a\otimes J^b\otimes P^c+\lambda  J^a\otimes P^b\otimes J^c),
\end{align}
where we introduced 
\bee 
\label{abrev}
\all  =\frac{\alpha^2+\lambda \beta^2}{(\tau\bar\tau)^2},\qquad 
\bett= -\frac{2\alpha\beta} {(\tau\bar\tau)^2},
\eee
so that
\begin{equation} \mu+\theta\nu=\frac{1}{\tau^2}. 
\end{equation}
In fact, the right-hand-side of \eqref{omegatau} is the  most general  general invariant element in $(\cg_\lambda)^3$.

We  sum up the discussion in the following definition.
\begin{definition} 
\label{compatible}
An $r$-matrix for any of the  Lie algebras $\cg_\lambda$ is compatible with the Chern-Simons action \eqref{CSaction} if $r=\rpr + K_\tau$, and $\rpr$ satisfies 
\bee
\label{mcybe}
[[\rpr ,\rpr ]] = - \Omega_\tau.
\eee
\end{definition} 

Our goal here is to find the most general  {\em real} solution of  equation \eqref{mcybe}.   At the level of  Lie algebras  we need to keep track of the real structures in order to  distinguish  between the various physically distinct regimes of 3d gravity. The reality of the classical $r$-matrices is required in the Fock-Rosly construction in order to have a real Poisson  structure  on the extended phase. This allows  for the usual interpretation of functions on the phase space as observables, and is expected  in the classical limit of any quantisation of the theory where observables are Hermitian with respect to a given $*$-structure.  

The combination of a  real Lie algebra structure with a  real $r$-matrix amounts, in the terminology of \cite{Majidbook},  to a real-real form of the Lie bialgebra structure defined by the $r$-matrix. It is mathematically possible to impose other conditions (for example the half-real structure defined in \cite{Majidbook} where the anti-symmetric part of the $r$-matrix is imaginary) but the relevance of such structures to  the generalised Chern-Simons formulation of 3d gravity  is not clear. We  therefore restrict our discussions  in the following to real $r$-matrices; the interested reader should have no difficulty in obtaining  solutions satisfying other reality conditions.

\section{Compatible $r$-matrices via generalised complexification}

In this section we show that a particular class of solutions of \eqref{mcybe} can be obtained by a process of (generalised) complexification of the standard solution of the MCYBE for the Lie algebra $sl(2,\RR)$. 
As explained in the appendix of the paper \cite{MSkappa}, the well-known identity
\begin{equation}
[X,[Y,Z]]=\langle X,Z \rangle Y-\langle X,Y \rangle Z\end{equation}
for the three dimensional Lie algebras $\cg$  can be used, together with the Jacobi identity and the invariance
of the Killing form, to establish that
 \begin{equation}
\rpr =\epsilon^{abc}m_{a}J_{b}\otimes J_{c}\end{equation}
satisfies\begin{equation}
[[\rpr ,\rpr ]]=m_{a}m^{a}\epsilon^{bcd}J_{b}\otimes J_{c}\otimes J_{d}.
\end{equation}
The quadratic Casimir\begin{equation}
K=J_{a}\otimes J^{a}\end{equation}
satisfies\begin{equation}
[[K,K]]=\Omega,\end{equation}
where $\Omega$ is the cubic Casimir $\Omega=\epsilon_{abc}J^{a}\otimes J^b\otimes J^{c}.$
Therefore, by \eqref{MCYBE}, 
\begin{equation}
r=K+\rpr \end{equation} satisfies the CYBE \eqref{CYBE} provided
\begin{equation}
m_{a}m^{a}=-1.\end{equation}
This has a non-trivial real solution only in the case $\cg =sl(2,\RR)$ and leads to the standard bialgebra structure in that case. 

Consider now the (generalised) complexification  $\cg_\lambda \otimes R_\lambda$ of the (real, 6-dimensional) Lie algebra $\cg_\lambda$.  The generators 
\begin{equation}
J_{a}^{\pm}=\frac{1}{2}(P_{a}\pm \theta J_{a})
\end{equation}
satisfy 
\begin{equation}
[J_{a}^{\pm},J_{b}^{\pm}]=\pm\theta\epsilon_{abc}(J^{\pm})^{c},\quad[J_{a}^{+},J_{b}^{-}]=0.
\end{equation}
It follows that 
\begin{equation}
K^{\pm}=J_{a}^{\pm}\otimes(J^{\pm})^{a}
\end{equation}
are both invariant element of the universal enveloping algebra $U(\cg_\lambda\otimes R_\lambda), $ and that
\begin{equation}
r^{\pm}(\tau,\bm)=\frac{1}{\tau}K^{\pm}+\epsilon^{abc}m_{a} J_{b}^{\pm}\otimes J_{c}^{\pm}
\end{equation}
both satisfy the CYBE over $R_{\lambda}$ for any invertible $\tau\in R_{\lambda}$ and vector $\bm \in R_{\lambda}^{3}$ satisfying the condition
\begin{equation}
\label{m^2=invtau}
\bm^2=-\frac{1}{\tau^{2}}.
\end{equation}
Since the $+$ and the $-$ copy of the Lie algebra $\cg_\lambda\otimes R_\lambda$ commute, we also deduce that any linear combination 
\begin{equation}
a^+r^+(\tau^+,\bm^+)+ a^-r^-(\tau^-,\bm^-)
\end{equation}
satisfies the CYBE for any $a^\pm\in R_\lambda,$ $\bm^\pm\in R^3_\lambda,$ provided  the condition \eqref{m^2=invtau}  holds for the parameters $\tau^\pm$ and $\bm^\pm.$
In particular, therefore,   the combinations
\begin{equation}
\label{combis}
r^+(\tau,\bm)\pm r^-(\bar{\tau},\bar{\bm})=r(\tau,\bm)\pm \overline{r(\tau,\bm)}\end{equation}
satisfy the CYBE. 

In order to  obtain a real solution of the CYBE for $\cg_\lambda$ with the symmetric part agreeing with the general Casimir element \eqref{Ktau},  we  note that 
\bee
 \frac 1 \tau K^+ - \frac {1} {\bar \tau} K^- =\frac {\theta }{2\tau\bar \tau}
\left(\alpha (P_a\otimes J^a  +J_a\otimes P^a)   - \beta (P_a\otimes P^a  +\lambda J_a\otimes J^a)\right).
\eee
Assuming for a moment  that $\lambda \neq 0$ (so that $\theta$ is no a zero-divisor), we can take the negative sign in \eqref{combis} and multiply the result  by $2/\theta$ to  obtain a solution with the  required symmetric part:
\bee
r(\tau,\bm)=\frac{2}{\theta} \left( r^+(\tau,\bm)-  r^-(\bar{\tau},\bar{\bm})\right).
\eee 
Writing 
\bee
\bm=\bp+\theta\bq, 
\eee
the solution takes the form
\bee
\label{complexr}
r(\tau,\bm)=K_{\tau}+\epsilon^{abc}(p_{a}(P_{b}\otimes J_{c}+J_{b}\otimes P_{c}) +q_{a}(P_{b}\otimes P_{c}+\lambda J_{b}\otimes J_{c})).
\eee
 Spelling out the condition for \eqref{complexr} to be a solution of the CYBE, we find
\begin{equation}
\label{condi}
\bp^{2}+\lambda\bq^{2}=-\mu,\quad 2 \bp\cdot \bq=-\nu ,
\end{equation}
where we used the abbreviations \eqref{abrev}.

The division by $\theta$ is potentially problematic in the case $\lambda=0$.  However, one can also check that \eqref{complexr} is a solution of the CYBE in 
the limit $\lambda\rightarrow0$  by making careful use of the identity
\begin{equation}
q_{a}p^{d}\epsilon_{dbc}+q_b p^{d}\epsilon_{adc}+q_{c}p^{d}\epsilon_{abd}=q_dp^d\epsilon_{abc}.
\end{equation}

One needs to check if  \eqref{condi} has solutions.  To simplify the analysis we define 
\bee
\bm'  = \tau \bm, 
\eee
and expand $
\bm'= \bp'+ \theta\bq'.$
The condition \eqref{m^2=invtau} is now simply
\bee
\bm'^2 = -1 
\eee
or 
\bee
(\bp ')^2 + \lambda (\bq')^2 = -1, \qquad \bp'\cdot \bq' =0.
\eee
In other words, solutions are determined by two orthogonal vectors whose squares satisfy  one constraint. 
When  $\lambda >0$, this constraint is clearly impossible to satisfy (for real vectors!) in the Euclidean case, but has various types of solutions in the Lorentzian case, including  $\bp'$ and $\bq'$ being orthogonal spacelike vectors, but also cases where only one of either $\bp'$ or $\bq'$ is spacelike and the other either lightlike or timelike. 

When $\lambda <0$, we can write 
the constraint as 
\bee
\label{negconst}
(\bp ')^2 + 1 = (- \lambda) (\bq')^2, \qquad \bp'\cdot \bq' =0.
\eee
This clearly has solutions in the Euclidean case. In fact, assuming $\lambda= -1$ without loss of generality, any such solutions may be interpreted as determining an ellipse in $\RR^3$ with minor axis $\bp'$ (including the degenerate case $\bp'=0$) and major axis $\bq'$.
There are also solutions in the Lorentzian case, of all the types described for the $\lambda >0 $ case above.

\section{Conditions for compatible $r$-matrices in 3d gravity}
\subsection{A Lie-algebraic equation}
We first  derive a Lie-algebraic condition for the most general solution of \eqref{mcybe}. In order to distinguish the various kinds of terms in that equation it is helpful to the use the notation of generalised complexification. Then   the invariant element \eqref{omegatau} takes the form
\begin{align}
\Omega_\tau &=(\all (\lambda \id\otimes \id \otimes \id + \id\otimes\theta \otimes \theta +
\theta\otimes\id\otimes \theta + \theta \otimes\theta \otimes \id ) + \bett( \theta \otimes \theta \otimes \theta \nonumber \\
& +\lambda \theta \otimes \id \otimes \id +\lambda \id \otimes \id \otimes \theta +\lambda  \id \otimes \theta \otimes \id))
\epsilon_{abc} J^a\otimes J^b\otimes J^c.
\end{align}
The antisymmetric part  $\rpr$ of the  $r$-matrix can be written as
\begin{align}
\label{guess}
\rpr &=(\id \otimes A + \theta  \otimes B - B\otimes \theta  + \theta \otimes \theta \;C)J^a\otimes J_a \nonumber \\
&=J^a\otimes A(J_a) + P^a\otimes B(J_a) - B(J_a)\otimes P^a + P^a\otimes C(P_a)\nonumber \\
&=A_{ba} J^a\otimes J^b + B_{ba} P^a\otimes J^b - B_{ba} J^b \otimes P^a + C_{ba} P^a\otimes P^b ,
\end{align}
where we can assume that $A$ and $C$ are anti-symmetric, i.e., $A_{ab}=-A_{ba}$ and $C_{ab}=-C_{ba}$.

Note that all the matrices appearing in \eqref{guess} should be thought of as matrices for linear maps 
\bee A,B,C: 
\cg\rightarrow \cg
\eee
 with respect to the orthonormal basis $\{J_0,J_1,J_2\}$ of $\cg$.
An antisymmetric map $A: \cg\rightarrow \cg$ can always and uniquely be expressed in terms of  an element $u=u_aJ^a \in \cg$, acting via commutator. Also note that
\bee
\label{Aurelation}
A= [u,\cdot] \;\;\Rightarrow \;\; A_{ab}= -\epsilon_{abc}u^c,
\eee
and that 
\bee
\label{ACrel}
 A= [u,\cdot],\; C= [v,\cdot]\;\; \Rightarrow \;\; AC=  vu ^t - \langle u,v \rangle \;\id,
\eee
which we will use frequently later in this paper.

Inserting \eqref{guess} into the left hand side of \eqref{mcybe} generates 48 terms. Setting them equal to the right hand side  of \eqref{mcybe} yields eight equations, each having six terms. We have found it useful to contract the equations with an element $X\otimes Y \otimes Z\in \cg^3$,  using the metric $\langle \cdot , \cdot \rangle$. That way, for example, the equation
\bee
[[J^a\otimes A(J_a), J^b\otimes A(J_b)]]= \epsilon_{abc}J^a\otimes J^b\otimes J^c
\eee
turns into
\bee
\langle X,[A^t(Y),A^t(Z)]\rangle + \langle Y,[A(X),A^t(Z)]\rangle + \langle Z,[A(X),A(Y)]\rangle  = \langle X, [Y,Z]\rangle  \quad \forall X,Y,Z \in \cg
\eee
which in turn is equivalent to
\bee
\label{socond}
 A([X,A^t(Y)]) +A ([Y,A(X)])+ [A(X),A(Y)] = [X, Y]\quad \forall X,Y \in \cg.
\eee
This is the sort of equation studied and solved  in \cite{OS2}. In the case at hand we can simplify the condition by using that, because of the antisymmetry $A^t=-A$,  $A$ is an adjoint action with a suitable Lie algebra element and therefore obeys the Jacobi identity
\bee
 A([X,Y])=[A(X),Y]+ [X,A(Y)].
\eee
Then the condition \eqref{socond} can be written as 
\bee
[A(X),A(Y)]- A^2([X,Y]) =[X, Y]\quad \forall X,Y \in \cg.
\eee

Proceeding similarly with  \eqref{mcybe} but omitting the explicit statement of $\forall X,Y\in\cg$,  we find, the following `raw' terms, where we have not yet used the antisymmetry of both $A$ and $C$.

From the $\id\otimes \id\otimes \id $ term:
\begin{align}
\label{111}
& A([X,A^t(Y)] + [Y,A(X)])+ [A(X),A(Y)]   \nonumber \\
 & \qquad  \qquad \qquad +\lambda(B([X,B^t(Y)] + [B^t(X),Y])+ [B^t(X),B^t(Y)]) = -\all\lambda[X, Y].
 \end{align}
From the $\theta\otimes \id\otimes \id$ term:
\begin{align}
\label{t11}
&   A([Y,B(X)]+[X,B^t(Y)])+ B([X,A^t(Y)])  +  [B(X),A(Y)]    \nonumber \\
 & \qquad  \qquad \qquad +\lambda(B([Y,C(X)]) + [B^t(Y),C(X)])= -\lambda \bett[X, Y].
 \end{align}
 From the $\id\otimes \theta\otimes \id$ term:
\begin{align}
\label{1t1}
&   B([Y,A(X)]) + [A(X),B(Y)] -A([X,B(Y)]+[Y,B^t(X)])    \nonumber \\
 & \qquad  \qquad \qquad +\lambda(B([X,C^t(Y)]) - [B^t(X),C(Y)])= -\lambda\bett[X, Y].
 \end{align}
 From the $\id \otimes \id\otimes \theta$ term:
\begin{align}
\label{11t}
& -B^t( [X,A^t(Y)] + [Y,A(X)]) -[B^t(X),A(Y)])-  [A(X),B^t(Y)]   \nonumber \\
 & \qquad  \qquad \qquad +\lambda(C([X,B^t(Y)])  - [Y,B^t(X)])= -\lambda \bett[X, Y].
 \end{align}
 From the $\theta\otimes \theta \otimes \id$ term:
\begin{align}
\label{tt1}
& B([Y,B(X)] - [X,B(Y)]) + [B(X),B(Y)]  
+A([X,C^t(Y)] +[Y,C(X)])  \nonumber \\
 & \qquad  \qquad \qquad +\lambda [C(X),C(Y)]) = -\all[X, Y].
 \end{align}
   From the $\theta\otimes \id \otimes \theta$ term:
\begin{align}
\label{t1t}
& -B^t([X,B^t(Y)] + [Y,B(X)])  -[B(X),B^t(Y)]  
+C([X,A^t(Y)] +[C(X),A(Y)])  \nonumber \\
 & \qquad  \qquad \qquad +\lambda C [Y,C(X)]) =- \all[X, Y].
 \end{align}
 From the $\id \otimes \theta \otimes \theta$ term:
\begin{align}
\label{1tt}
& B^t([X,B(Y)]) +[Y,B^t(X)])  -[B^t(X),B(Y)]  
+C([Y,A(X)] +[A(X),C(Y)])  \nonumber \\
 & \qquad  \qquad \qquad +\lambda C [X,C^t(Y)]) = -\all[X, Y].
 \end{align}
  Finally, from the $\theta \otimes \theta \otimes \theta$ term:
\begin{align}
\label{ttt}
& -B^t([X,C^t(Y)]) -B^t [Y,C(X)])  +[B(X),C(Y)]  +[C(X),B(Y)]
 \nonumber \\
 & \qquad  \qquad \qquad +C([Y,B(X)]  - [X,B(Y)])  = - \bett [X, Y].
 \end{align}

We can simplify the equations \eqref{111} and \eqref{ttt} using the antisymmetry of $A$ and $C$ to find
\begin{align}
\label{111m}
 [A(X),A(Y)] -A^2([X,Y])+\lambda(B([X,B^t(Y)] + [B^t(X),Y])+[B^t(X),B^t(Y)]) = -\all\lambda[X, Y].
 \end{align}
and 
\begin{align}
\label{tttm}
 B^t C([X,Y])  +[B(X),C(Y)]  +[C(X),B(Y)]
 -C([B(X),Y] + [X,B(Y)])  = - \bett [X, Y].
 \end{align}

The equations \eqref{tt1}-\eqref{1tt} turn out to be mutually equivalent. The best way to see this is to contract again with a general vector $Z$ and to use cyclic identities. Thus we can replace the three equations by the single equation obtained from \eqref{tt1} after using the antisymmetry of $C$ and the Jacobi identity:
\begin{align}
\label{tt1m}
&  [B(X),B(Y)]- B([X,B(Y)] +[X,B(Y)]) 
-AC([X,Y])  +\lambda [C(X),C(Y)] = -\all[X, Y].
 \end{align}

The equations \eqref{t11}-\eqref{11t} are also mutually equivalent. We can see this again by projecting onto $Z$, using cyclic identities and re-naming. Using also the antisymmetry of $A$ and $C$, we obtain
\begin{align}
\label{11tmm}
B^t A( [X,Y]) -[B^t(X),A(Y)])-  [A(X),B^t(Y)]  +\lambda C([X,B^t(Y)] + [B^t(X),Y])= -\lambda \bett[X, Y].
 \end{align}

We thus obtain a  set of four coupled equations for the  linear maps  $A,B,C$. We combine them here for clarity:
\begin{align}
\label{master}
{}[A(X),A(Y)]-A^2([X,Y])+\lambda(B([X,B^t(Y)] + [B^t(X),Y])+ [B^t(X),B^t(Y)]) &= -\all\lambda[X, Y], \quad  \nonumber \\
B^t C([X,Y])  +[B(X),C(Y)]  +[C(X),B(Y)]
 -C([B(X),Y] + [X,B(Y)]) & =  -\bett [X, Y],\quad   \nonumber \\
[B(X),B(Y)]- B([X,B(Y)] +[X,B(Y)]) 
-AC([X,Y])  +\lambda [C(X),C(Y)] &=- \all[X, Y],\quad   \nonumber \\
B^t A( [X,Y]) -[B^t(X),A(Y)])-  [A(X),B^t(Y)]  +\lambda C([X,B^t(Y)] + [B^t(X),Y])&=- \lambda \bett[X, Y].\quad  \nonumber \\
\qquad \qquad\qquad\forall X,Y\in \cg\qquad\qquad &
 \end{align}

\subsection{An equation involving three linear maps}

Our main tool in this section is  Lemma 3.1 of \cite{OS2}. We recall it here for the convenience of the reader. 
\begin{lemma}
For every linear map $F: \cg\rightarrow \cg$, there is a uniquely determined linear map
\bee
F^\Ad:\cg \rightarrow \cg, 
\eee
which satisfies
\bee
\label{defad}
\langle F^\Ad (Z),[X,Y]\rangle = \langle Z,[F(X),F(Y)]\rangle \quad \forall X,Y,Z \in \cg.  
\eee
It is  given by
\bee
\label{Fmaster}
F^\Ad= F^2-\tr(F)\, F +\frac 1 2 \left (\tr(F)^2-\tr(F^2)\right)\id ,
\eee
which is the adjugate of $F$. 
\end{lemma}
We need a corollary of this lemma, which is obtained by polarisation:
\begin{lemma}
For  linear maps $E,F: \cg\rightarrow \cg$, there is a uniquely determined linear map
\bee
(E,F)^\Ad:\cg \rightarrow \cg, 
\eee
which satisfies
\bee
\label{genad}
\langle (E, F)^\Ad (Z),[X,Y]\rangle = \langle Z,[(E(X),F(Y)]+ [F(X),E(Y)]\rangle \quad \forall X,Y,Z \in \cg.  
\eee
It is  given by
\bee
\label{EFmaster}
(E, F)^\Ad= FE+EF-\tr(F)\,E -\tr(E)\, F + (\tr(F)\tr(E)-\tr(EF))\id.
\eee 
\end{lemma}

It follows that 
\bee
(\id, B)^\Ad= -B+\tr(B)\,\id
\eee

Thus,  taking adjugates of equation \eqref{111m},  we obtain, using that $\tr(A)=0$,  
\begin{align}
&-\frac 1 2 \tr(A^2)\id +\lambda( (-B^t+\tr(B)\,\id) B^t\nonumber \\ &\qquad \qquad \qquad  + (B^t)^2 -\tr(B)B^t +\frac 1 2 \left  (\tr(B)^2-\tr(B^2)\right)\id =-\all \lambda\, \id,
\end{align}
or 
\begin{align}
-\frac 1 2 \tr(A^2)\id +\frac \lambda 2 \left  (\tr(B)^2-\tr(B^2)\right)\id =-\all \lambda \, \id.
\end{align}

Similarly taking adjugates of equation \eqref{tttm} and using $\tr (C)=0$,  we find 
\begin{align}
 -CB+(BC+CB-\tr(B)\,C -\tr(CB)\id) +   ( -B+\tr(B)\,\id   )C=-\bett \, \id, 
\end{align}
or 
\begin{align}
 - \tr(CB)\id =-\bett \id, 
\end{align}

Proceeding with the equation \eqref{tt1m} in a similar fashion gives
\begin{align}
& B^2-\tr(B)\, B+\frac 1 2 \left (\tr(B)^2-\tr(B^2)\right)\id - (-B+\tr(B)\,\id)B^t-CA \nonumber \\ 
& \qquad \qquad \qquad\qquad \qquad \qquad  +\lambda  (C^2 -\frac 1 2 \tr(C^2)\id)
=-\all \, \id 
\end{align}
or 
\begin{align}
 (B-\tr(B))(B+B^t)+\frac 1 2 \left (\tr(B)^2-\tr(B^2)\right)\id-CA +\lambda  (C^2 -\frac 1 2 \tr(C^2)\id)=-\all \,  \id 
\end{align}

Finally, we obtain from equation \eqref{11tmm} that 
\begin{align}
 -AB - (AB^t+B^tA-\tr(B)\,A  -\tr(AB^t)\id) -\lambda    ( -B^t+\tr(B)\,\id ) C =-\lambda \bett\, \id  
\end{align}
or 
\begin{align}
 -A(B+B^t) - (B^t -\tr(B))\,A  -\tr(AB)\id +\lambda    ( B^tC -\tr(B)C) =-\lambda \bett\, \id  
\end{align}

The conclusion  from these calculations is  one of the main results of this paper:
\begin{theorem}
The $r$-matrix $r=\rpr+K_\tau$ with $\rpr$ given in \eqref{guess} is compatible with the Chern-Simons action  \eqref{CSaction}  in the sense of definition \ref{compatible} 
if the linear maps $A,B$ and $C$  satisfy the  following coupled equations:
\begin{align}
 \label{masterr}
 \dfrac 1 2 \tr(A^2) -\dfrac \lambda 2 \left  (\tr(B)^2-\tr(B^2)\right) &=\all \lambda ,\nonumber \\
\tr(CB) &=\bett, \nonumber\\
(B-\tr(B)\id)(B+B^t)+\dfrac 1 2 \left (\tr(B)^2-\tr(B^2)\right)\id-CA +\lambda  (C^2 -\dfrac 1 2 \tr(C^2)\id)&=-\all\, \id, \nonumber \\
-A(B+B^t) + (B^t -\tr(B)\id)\,(\lambda C- A)  -\tr(AB)\id  &=-\lambda \bett  \, \id. 
\end{align}
\end{theorem}

The coupled matrix  equations \eqref{masterr} are non-linear but can be analysed with  linear algebra of the sort used in \cite{OS2} and \cite{Stachura}. We have not been able  obtain a  complete set of solutions. In the  next  section, we give two classes of solutions which contain some  compatible $r$-matrices known in the literature, as well as  new ones.

\section{Two families of compatible 
$r$-matrices}

\subsection{Solutions for the case $A=\lambda C$}
It is possible to determine a family of  solutions by making the ansatz $A=\lambda C$. If  $\lambda\neq 0$, the equations  (\ref{masterr})  reduce to
\bea
\frac {\lambda}{ 2} \tr(C^2) -\frac 1 2 \left  (\tr(B)^2-\tr(B^2)\right) &= \all, \label{BBsimp} \\
\tr(CB)& =\bett,\label{CBsimp} \\
(B-\tr(B)\id)(B+B^t)&=0,\label{BBtB}\\
-\lambda C(B+B^t) &=0. \label{BBtsimp}
\eea
We can solve this system in terms of  non-zero element $
v \in \cg $, a further element $w\in \cg$ 
and  real numbers $x,y$ by parametrising (without loss of generality) 
\bee
C=x [v, \cdot],
\eee
and making the further ansatz
\bee
\label{Bansatz}
B= y\,  v v^t +  [w,\cdot ].
\eee
Then  $\tr (B) = y \langle v, v \rangle$  and 
\bee
B+B^t = 2 y\;  v v^t \qquad B-\tr(B)\id = y ( v v^t - \langle v, v \rangle\id )  +  [w,\cdot ],
\eee
so that \eqref{BBtB} and  \eqref{BBtsimp} are  satisfied
provided 
\bee
\label{crux}
y[v,w]=0.
\eee
It follows that 
\bee 
B^2=  y^2\langle v, v \rangle v v^t + ww^t   - \langle w, w \rangle  \id, 
\eee
Moreover, if $y\neq 0$, the condition \eqref{crux} means 
\bee
w=zv, \quad z \in \RR.
\eee
Assuming this, 
one checks that 
\bee
\frac 1 2 \left  (\tr(B)^2-\tr(B^2)\right)= z^2 \langle v,v \rangle.
\eee
so that  \eqref{BBsimp} and \eqref{CBsimp} give us the two normalisation conditions 
\bee
\label{ACcondition}
-(\lambda x^2 + z^2)\langle v,v \rangle= \mu, \qquad -2xz\langle v,v \rangle = \nu.
\eee

A particular solution in this case is the `special' double solutions in \cite{BHM}. In this case $\mu >0$, $ \lambda >0$, the metric is necessarily Lorentzian and the solution is  parametrised by a spacelike  element $v\in \cg$ satisfying $\langle v, v \rangle  =-1$.
Comparing with  equation 5.26 in  \cite{BHM}, the parameter $\rho$ used there is related to our parameters $x,y,z$ via
\bee
x=\frac{1-\rho^2}{4\sqrt{\lambda}}, \qquad y = \frac \rho 2, \qquad z= \frac{1+\rho^2}{4}, 
\eee
so that 
\bee
\lambda x^2 +  z^2 =\frac{1+\rho^4}{8} ,\qquad 
2xz = \frac{1-\rho^4}{8\sqrt{\lambda}}.
\eee
From equation 5.25 in  \cite{BHM},  our parameters $\all,\bett$ can be written in terms of the parameter $\rho$ is as
\bee
\all =\frac{1+\rho^4}{8} , \qquad \bett = \frac{1-\rho^4}{8\sqrt{\lambda}},
\eee
so that, with $\langle v, v \rangle  =-1$, our condition \eqref{ACcondition} is indeed satisfied.

To sum up, the most general solution with $\lambda \neq 0$, $A=\lambda C$ and $B$ of the form \eqref{Bansatz} with  $y\neq 0$ has an $r$-matrix with anti-symmetric part 
\bee
\label{genACsol}
r' = (y v^av^b + zv^c\epsilon_{abc}) (P^a\otimes J^b -  J^b \otimes P^a) + x v^c\epsilon_{abc}(\lambda J^a\otimes J^b + P^a\otimes P^b) ,
\eee
and $x,z$ and $v$ satisfying \eqref{ACcondition}. The real parameter $y$ can take any value.

If $y=0$,  we need no longer require that $[v,w]=0$ and then \eqref{genACsol} is not the most general solution of the form \eqref{Bansatz}.  In this case,  the family \eqref{genACsol} coincides with  the family of solutions obtained by complexification in Sect.~3. To make contact with the notation there,  we expand
\bee
B_{ab} = - \epsilon_{abc} p^c, \qquad C_{ab} =-\epsilon_{abc}  q^c,\qquad A_{ab}=\lambda C_{ab} .
\eee
Then 
\bee
 \tr(B)=0, \quad \tr(B^2)= -2\bp^2, \quad \tr(C^2) =-2 \bq^2, \quad \tr (B C) = -2 \bp\cdot \bq, 
 \eee
  and so \eqref{BBsimp} and \eqref{CBsimp} become 
the condition  \eqref{condi} derived via complexification.

Finally, we turn to the  limiting case $\lambda=0$. In that case, the ansatz $A=\lambda C$ means that $A=0$, and the first and fourth  equation in \eqref{masterr} are automatically satisfied. In order to retain the simplification in the third equation we need to impose
\bee
 -\frac 1 2 \left  (\tr(B)^2-\tr(B^2)\right) = \all,
\eee
and solve \eqref{BBtB}. The matrix $C$ can be chosen freely subject to the constraint \eqref{CBsimp}.
This reproduces the $\lambda=0$ solutions discussed in Sect.~3, but also includes solutions of the form \eqref{genACsol} with $y\neq 0$. 

\subsection{ Solution for $\bett=0$  }

It is important to classify solutions compatible with either  the  gravitational pairing  \eqref{pair} or the  pairing \eqref{othform}.  This case is characterised by $\alpha \beta=0$ or $\bett=0$.  We obtain the complete family of solutions and  highlight the new ones. 

We first show that, if $\bett =0$, and  $\lambda \neq 0$, one can use the  fourth equation in \eqref{masterr}  to express $B$ in terms of $A$ and $C$.
Inserting the ansatz
\bee
\label{BAC}
B=AC + x\; \id
\eee
into the fourth equation, and again writing the antisymmetric maps $A$ and $B$ in terms of Lie algebra elements $u,v$ as 
\bee
\label{ACuv}
A=[u,\cdot], \qquad C=[v,\cdot],
\eee
we can use \eqref{ACrel} to write 
\bee
\label{Buv}
B= vu^t + (x-\langle u, v\rangle)\,\id
\eee
and hence to write the equations \eqref{masterr}  in terms of $u,v$ and $x$. In particular, one finds 
\bee
\frac{1}{2} (\tr (B)^2 -\tr(B^2)) = (2x-\langle u, v\rangle)^2 -x^2.
\eee
Inserting\eqref{ACuv} and  \eqref{Buv} in the fourth equation of \eqref{masterr},  and setting $\bett=0$  leads to 
\bee
\label{uv4}
-\langle u, w\rangle[u,v]  + \langle u, v\rangle  [u,w]
-\langle v,[u,w] \rangle u +\lambda  (\langle u, v\rangle - 2x)[v,w]  =0.
\eee
It follows from the  identity
\begin{align}
[w,[u,[u,v]]] &= \langle u,v \rangle [w,u] - \langle u,u \rangle [w,v]
= \langle w, [u,v] \rangle   u  - \langle w, u \rangle [u,v],
\end{align}
that 
\bee
\langle u,v \rangle [w,u]+\langle w, u \rangle [u,v] +\langle w, [v,u] \rangle   u  = \langle u,u \rangle [w,v].
\eee
Using this to simplify \eqref{uv4}, one concludes that, for $\lambda \neq 0$,  the  fourth equation in \eqref{masterr}  is equivalent to
\bee
\label{xB}
x= \frac 1 2 \left(\langle u, v\rangle + \frac{\langle u, u\rangle }{\lambda}\right).
\eee

Inserting \eqref{Buv} with this expression for $x$ into the third equation in \eqref{masterr},  yields
\bee
(\langle u, u,\rangle +\lambda)(v v^t - \frac 1 \lambda  u v^t)-\frac 1 4 \left(\langle u,v \rangle -\frac{\langle u, u\rangle }{\lambda}\right)^2\id  + \langle u,v\rangle \id =-\mu\, \id.
\eee
This is equivalent to 
\bee
\label{nucond}
\langle u , u \rangle = -\lambda, \qquad (\langle u ,v \rangle -1)^2 =4\mu.  
\eee

The second equation in \eqref{masterr} is automatically satisfied for $B$ of the form \eqref{BAC} since $CAC$ is antisymmetric and so $\tr(CAC)=0$. Finally, if  $\lambda = -\langle u , u \rangle$  then $x= \frac 12 (\langle u,v\rangle -1)$ and so the first equation in \eqref{masterr} becomes 
\bee
\lambda - \lambda (1-x^2) = \lambda \mu,
\eee
which, for $\lambda \neq 0$, is equivalent to  the second  equation in \eqref{nucond}.

The equation \eqref{nucond} has real solutions when $\mu  \geq 0$. When  $ \lambda <0$ there are solutions for both Euclidean and Lorentzian signatures, but when $\lambda >0$ we require Lorentzian signature and a spacelike element $u$. Assuming these conditions are met, the only condition on $C$ is that 
\bee
\langle u , v \rangle = 1 \pm 2\sqrt{\mu}.
\eee
The  map $B$ then takes the form
\bee
 B= vu^t -(1\pm \sqrt{\mu})\,\id,
\eee
and the antisymmetric part of the $r$-matrix is 
\bee
\label{genbett0sol}
r' = \left( u^av^b- (1\pm \sqrt{\mu})\eta_{ab}\right) (P^a\otimes J^b - J^b \otimes P^a) +  u^c\epsilon_{abc} J^a\otimes J^b + v^c\epsilon_{abc}P^a\otimes P^b, 
\eee
with $\lambda \neq 0$ and  $u,v \in \cg$ subject to   the constraints \eqref{nucond}.

\section{Conclusion}

The set of equations \eqref{masterr} determines the most general $r$-matrix which is compatible with the generalised Chern-Simons action for 3d gravity. 
In Table~\ref{summary}  we summarise the solutions found in this paper. Some of the solutions listed there, like the `standard doubles' and `generalised bicrossproducts' for some parameter values, have been known for some time, but the family of solutions obtained  by generalised complexification and the family of solutions for $\bett=0$ are new.

\begin{table}[h]
\centering
\begin{tabular}{|c|c|c|c|}
\hline 
&& \\
Type  & Solution & Constraint  \\
&&\\
\hline
&&\\
Standard doubles & $B=\pm \sqrt{\mu} \,\id $, $A=[p,\cdot] $, $C=0$  &$\all \geq 0$,  $\langle p , p \rangle = -\lambda$ \\
&&\\
 Generalised bicrossproducts &  $B=y\, vv^t+z[v,\cdot]$,  &$ (\lambda x^2 + z^2)\langle v,v \rangle= -\mu$, \;\;   \\
 &$A=\lambda C$, $C=x\,[v,\cdot]$ &$ 2xz\langle v,v \rangle = -\nu$ \\
 &&\\
Generalised complexifications & $A=\lambda C$, $B=[p,\cdot]$,  $ C=[q,\cdot]$& $\langle p , p \rangle +\lambda\langle q , q \rangle =-\mu,\;\; $  \\
  & &  $2 \langle p , q \rangle =-\nu$\\
&& \\
 Solutions for $\bett=0$ & $A=[u,\cdot]$,  $B=[v,\cdot] $, & $\mu \geq 0$, $\langle u , u \rangle = -\lambda, $  \\
     &$ B= vu^t -(1\pm \sqrt{\mu})\, \id $   & $(\langle u ,v \rangle -1)^2 =4\mu$ \\
  && \\   
\hline
\end{tabular}
 \vspace{.5cm}
\caption{Types of compatible $r$-matrices discussed in this paper: the maps $A,B,C$ parametrise $\rpr$ via   equation \eqref{guess} so that $\rpr + K_\tau$, with $K_\tau$ given in \eqref{Ktau} satisfies the classical Yang-Baxter equation \eqref{CYBE}. The names  are not standard; note in particular that some of the `generalised bicrossproducts' may also be viewed as doubles for certain parameter values, see the discussion in Sect.~5.1 and in   \cite{BHM}.}
\label{summary}
 \end{table}

 It remains a challenge to determine all solutions systematically. 
Meeting this challenge would  allow us to understand possible non-commutative geometries arising in the quantisation of  (generalised) 3d gravity and the relation between them.

It would also be of interest in the context of gravitational scattering. 
 When studying 3d gravity in a universe with compact spatial slices, one expects  different $r$-matrices compatible with the same Chern-Simons action to give rise to equivalent quantum theories. However, when the spatial slices have boundaries,  different $r$-matrices may encode different physics. In \cite{BaisMullerSchroers}, for example,  it was shown how  quantum $R$-matrices determine scattering of massive particles in a universe with non-compact spatial slices. Corresponding results for other scattering processes, for example of massive particles with BTZ black holes in AdS$_3$,  are not known, but  would require quantum $R$-matrices corresponding  the solutions of our set of equations \eqref{masterr}.

\section*{Acknowledgements}
PKO thanks the Perimeter Institute and the Fields Institute for the Fields-Perimeter Africa Postdoctoral Fellowship, and  the University of Ghana for its support. BJS thanks the Perimeter Institute for hosting a research visit. 
This research was supported in part by the  Perimeter Institute for Theoretical Physics. Research at Perimeter Institute is supported by the Government of Canada through the Department of Innovation, Science and Economic Development Canada and by the Province of Ontario through the Ministry of Research, Innovation and Science.


\begin{thebibliography}{99}


\bibitem{AGSI} A.~Y.~Alekseev, H.~Grosse and V.~Schomerus, 
Combinatorial quantization of the Hamiltonian Chern-Simons Theory,
Commun.~Math.~Phys.  172 (1995) 317--358.

\bibitem{AGSII} A.~Yu.~Alekseev, H.~Grosse and V.~Schomerus, 
Combinatorial quantization of the Hamiltonian Chern-Simons Theory
II, Commun.~Math.~Phys.  174 (1995) 561--604.

\bibitem{AS} A.~Yu.~Alekseev and V.~Schomerus,  Representation
theory of Chern-Simons observables, Duke Math.~Journal 85
(1996) 447--510.



\bibitem{Schroers} B.~J.~Schroers, Combinatorial quantisation of
Euclidean gravity in three  dimensions, in:  N.~P.~Landsman,
M.~Pflaum, M.~Schlichenmaier (Eds.), {
Quantization of singular symplectic quotients}, 
Progress in Mathematics, Vol.~198,
 2001, 307--328; { math.qa/0006228}.

\bibitem{BNR} E.~Buffenoir, K.~Noui and P.~Roche,  Hamiltonian
Quantization of Chern-Simons theory with $SL(2,\CC)$ Group,
Class.~Quant.~Grav. 19 (2002)  4953-5016.

\bibitem{MeusburgerSchroers1} C.~Meusburger and  B.~J.~Schroers, Poisson structure and
  symmetry in the Chern-Simons formulation of (2+1)-dimensional
  gravity,
Class.~Quant.~Grav. 20 (2003) 2193--2233.

\bibitem{MeusburgerSchroers2} C.~Meusburger and  B.~J.~Schroers, The quantisation of Poisson
    structures arising in Chern-Simons theory with gauge group
    $G\ltimes \mathfrak{g}^*$, Adv.~Theor.~Math.~Phys.~7 (2004)
    1003--043.

\bibitem{MN} C. Meusburger and K. Noui, The Hilbert space of 3d gravity: quantum group symmetries and 
observables, Adv.~Theor.~Math.~Phys.~14, 6 (2010) 1651--1716.

\bibitem{AT} A.~Achucarro and P.~Townsend, A Chern--Simons
 action for three-dimensional anti-de Sitter supergravity
 theories, Phys.~Lett. B  180 (1986) 85--100.

\bibitem{Witten} E.~Witten, 2+1 dimensional gravity as an exactly
 soluble system,Nucl.~Phys. B {311} (1988)  46--78.


\bibitem{Carlipbook} S.~Carlip,  Quantum gravity in 2+1 dimensions,
Cambridge University Press, Cambridge, 1998.


\bibitem{FockRosly} V.~V.~Fock and  A.~A.~Rosly, Poisson structures on
moduli of flat connections on Riemann surfaces and $r$-matrices,
ITEP preprint (1992)72--92; see also math.QA/9802054.

\bibitem{sissatalk} B.~J.~Schroers,  Lessons from (2+1)-dimensional quantum gravity,  Proceedings PoS (QG-Ph) 035 
''From Quantum to Emergent Gravity: Theory and Phenomenology``,  Trieste 2007; see also arXiv:0710.5844 [gr-qc].


\bibitem{Cracow} B.~J.~Schroers, Quantum gravity and non-commutative spacetimes in three dimensions: a unified approach,  Acta Phys. Pol. B Proceedings Supplement vol. 4 (2011)  379. 


\bibitem{MSkappa} C.~Meusburger and B.~J.~Schroers, Generalised Chern-Simons actions for 3d gravity and kappa-Poincare symmetry, Nucl. Phys. B 806 (2009) 462--488.

\bibitem{OseiSchroers1} P. ~K.~Osei and B.~J.~Schroers, On Semiduals of local isometry groups in 3d gravity, J.~Math.~Phys.~53  (2012) 073510.


\bibitem{BHM} A.~Ballesteros, F.~J.~Herranz, C.~Meusburger
Drinfel'd doubles for (2+1)-gravity, Class.~Quant.~Grav.~30 (2013) 155012. 


\bibitem{MielkeBaekler} E. ~W.~Mielke and P. ~Baekler, Topological gauge model of gravity with torsion, Phys. Lett. A 156 (1991) 399--403.


\bibitem{BonzomLivine} V.~Bonzom and E.~R.~Livine, A Immirzi-like parameter for
3d quantum gravity, Class.~Quant.~Grav. 25 (2008) 195024.

\bibitem{AGKY} J. B. Achour, M. Geiller, K. Noui, and C. Yu, Testing the role of the Barbero-Immirzi parameter and the choice of connection in Loop Quantum Gravity, Phys.~Rev.~D91 (2015)  104016.

\bibitem{Stachura} P. Stachura, Poisson-Lie structures on Poincar\'e and Euclidean groups in three dimensions, J.~Phys.~A:~Math.~Gen. 31 (1998) 4555--4564.

\bibitem{OS2}P.~K.~Osei and B.~J.~Schroers, Classical r-matrices via semidualisation, J.~Math.~Phys.~54 (2013) 101702. 

\bibitem{Mess}G.~Mess,  Lorentz spacetimes of constant curvature, preprint IHES/M/90/28,  1990. 

\bibitem{Messcomment} L.~Andersson, T.~Barbot, R.~Benedetti, F.~Bonsante, W.~M. ~Goldman, F.~Labourie, K.~P. ~Scannell and J.~-M.~Schlenker,
Notes on a paper of Mess, Geometriae Dedicata 126 (2007) 47--70, see also arXiv:0706.0640.

\bibitem{Matschull}  H.~-J.~Matschull,
{On the relation between (2+1) Einstein gravity and Chern-Simons
theory}, { Class.~Quant.~Grav.} { 16} (1999) 2599--2609.


\bibitem{TM} R.~Tresguerres and E.~W.~Mielke,
 Gravitational Goldstone fields from affine gauge theory,  Phys.~Rev.~D  62 (2000) 044004.

\bibitem{Meusburger} C.~Meusburger,  Geometrical (2+1)-gravity and the
Chern-Simons formulation: Grafting, Dehn twists, Wilson loop
observables and the cosmological constant, Commun.~Math.~Phys.~273 (2007) 705--754.

\bibitem{MSquart} C.~Meusburger and B.~J.~Schroers,   Quaternionic and Poisson-Lie structures in 3d gravity: the cosmological constant as deformation
parameter, J.~Math.~Phys.~49 (2008) 083510.


\bibitem{CP} V.~Chari and A.~Pressley, A guide to quantum groups, Cambridge University Press, Cambridge, 1994.

 \bibitem{Majidbook} S.~Majid, Foundations of quantum
group theory , Cambridge University Press, Cambridge 1995. 


\bibitem{BaisMullerSchroers}
  F.~A.~Bais, N.~M.~Muller, B.~J.~Schroers, Quantum  group symmetry and particle scattering in (2+1)-dimensional quantum gravity, Nucl.~Phys.  B640 (2002) 3--45.


 

\end{thebibliography}
\end{document}